\documentclass{www2012-comp-accepted}
\usepackage[tight,footnotesize]{subfigure}
\usepackage{cite}
\usepackage{multirow}
\begin{document}
\conferenceinfo{WWW 2012 Companion,} {April 16--20, 2012, Lyon, France.} 
\CopyrightYear{2012} 
\crdata{978-1-4503-1230-1/12/04} 
\clubpenalty=10000 
\widowpenalty = 10000

\title{Improving News Ranking by Community Tweets}
%Format\titlenote{(Does NOT produce the permission block, copyright information nor page numbering). For use with ACM\_PROC\_ARTICLE-SP.CLS. Supported by ACM.}}
%\subtitle{[Extended Abstract]
%\titlenote{A full version of this paper is available as
%\textit{Author's Guide to Preparing ACM SIG Proceedings Using
%\LaTeX$2_\epsilon$\ and BibTeX} at

\numberofauthors{3} %  in this sample file, there are a *total*
\author{
%Xin Shuai$^\star$, Ying Ding$^\ddagger$,Jerome Busemeyer$^\diamond$,Shanshan Chen$^\ddagger$,Yuyin Sun$^\ddagger$,Jie Tang$^\sharp$\\
%$^\star$School of Informatics and Computing, Indiana University Bloomington\\
%$^\ddagger$School of Library and Information Science, Indiana University Bloomington\\
%$^\diamond$Department of Psychological and Brain Science, Indiana University Bloomington\\
%$^\sharp$Department of Computer Science and Technology, Tsinghua University\\
%\email{\{xshuai, dingying, chenshan, yuysun, jbusemey\}@indiana.edu}, \email{jery.tang@gmail.com}
%Alberto Pepe$^{2}$,
%Johan Bollen$^{1}$
%\\
%\bf{1} School of Informatics and Computing, Indiana University, Bloomington, IN
%\\
%\bf{2} Center for Astrophysics, Harvard University, Cambridge, MA
%\\
%1st. author
%\alignauthor
Xin Shuai\\
       \affaddr{School of Informatics and}\\ 
       \affaddr{Computing}\\
       \affaddr{Indiana University}\\
       \affaddr{Bloomington}\\
       \affaddr{IN, USA}\\
       \email{xshuai@indiana.edu}
% 2nd. author
\alignauthor
Xiaozhong Liu\\
       \affaddr{School of Library and Information Science}\\
       \affaddr{Indiana University Bloomington}\\
       \affaddr{IN, USA}\\
       \email{liu237@indiana.edu}
% 3rd. author
\alignauthor 
Johan Bollen\\
       \affaddr{School of Informatics and Computing}\\
       \affaddr{ Indiana University Bloomington}\\
       \affaddr{IN, USA}\\
       \email{jbollen@indiana.edu}
%\and  % use '\and' if you need 'another row' of author names
% 4th. author
%\alignauthor 
%Shanshan Chen\\
%      \affaddr{School of Library and Information Science}\\
%       \affaddr{Indiana University Bloomington}\\
%       \affaddr{IN, USA}\\
%       \email{chenshan@indiana.edu}
%% 5th. author
%\alignauthor
% Yuyin Sun\\
%      \affaddr{School of Library and Information Science}\\
%       \affaddr{Indiana University Bloomington}\\
%       \affaddr{IN, USA}\\
%       \email{yuysun@indiana.edu}
%% 6th. author
%\alignauthor 
%Jie Tang\\
%      \affaddr{Dept. of Computer Science and Technology}\\
%       \affaddr{Tsinghua University}\\
%       \affaddr{Beijing, China}\\
%       \email{jery.tang@gmail.com}
}
% There's nothing stopping you putting the seventh, eighth, etc.
% author on the opening page (as the 'third row') but we ask,
% for aesthetic reasons that you place these 'additional authors'
% in the \additional authors block, viz.
%\additionalauthors{Additional authors: John Smith (The Th{\o}rv{\"a}ld Group,
%email: {\texttt{jsmith@affiliation.org}}) and Julius P.~Kumquat
%(The Kumquat Consortium, email: {\texttt{jpkumquat@consortium.net}}).}
% Just remember to make sure that the TOTAL number of authors
% is the number that will appear on the first page PLUS the
% number that will appear in the \additionalauthors section.

\maketitle
\begin{abstract}
Users frequently express their information needs by means of short and general queries that are difficult for ranking algorithms to interpret correctly. However, users' social contexts can offer important additional information about their information needs which can be leveraged by ranking algorithms to provide augmented, personalized results. Existing methods mostly rely on users' individual behavioral data such as clickstream and log data, but as a result suffer from data sparsity and privacy issues. Here, we propose a Community Tweets Voting Model (CTVM) to re-rank Google and Yahoo news search results on the basis of open, large-scale Twitter community data. Experimental results show that CTVM outperforms baseline rankings from Google and Yahoo for certain online communities. We propose an application scenario of CTVM and provide an agenda for further research. 
\end{abstract}

\category{H.3.3}{Information Storage and Retrieval}{Information Search and Retrieval}

%A category including the fourth, optional field follows...
%\category{D.2.8}{Software Engineering}{Metrics}[complexity measures, performance measures]

%\terms{Algorithms, Experimentation}

\keywords{Twitter, news ranking, community interest} % NOT required for Proceedings

\section{Introduction}
Many news search engines, like Google and Yahoo, allow users to search across thousands of news sources with a single search query. Given the scale of online information, any given query can match vast numbers of news results. Search engines therefore use ranking mechanisms to prioritize search results to favor those that are estimated to be most relevant to users. The quality of their rankings has therefore become an important criterion to measure the performance of a news search engine.

Unfortunately, existing topology-based ranking algorithms, like PageRank or HITs, may not be appropriate for news ranking. News information is frequently not well-embedded in the hyperlink topology of the web. In addition, it is by definition highly dynamic, and designed to respond to rapidly changing user preferences. An effective news search engine is thus charged with providing personalized ranking results that are not solely based on document relevance and hyperlink connections, but also take into account dynamic user information needs.

\begin{figure}[htbp]\centering
	\includegraphics[width=6cm]{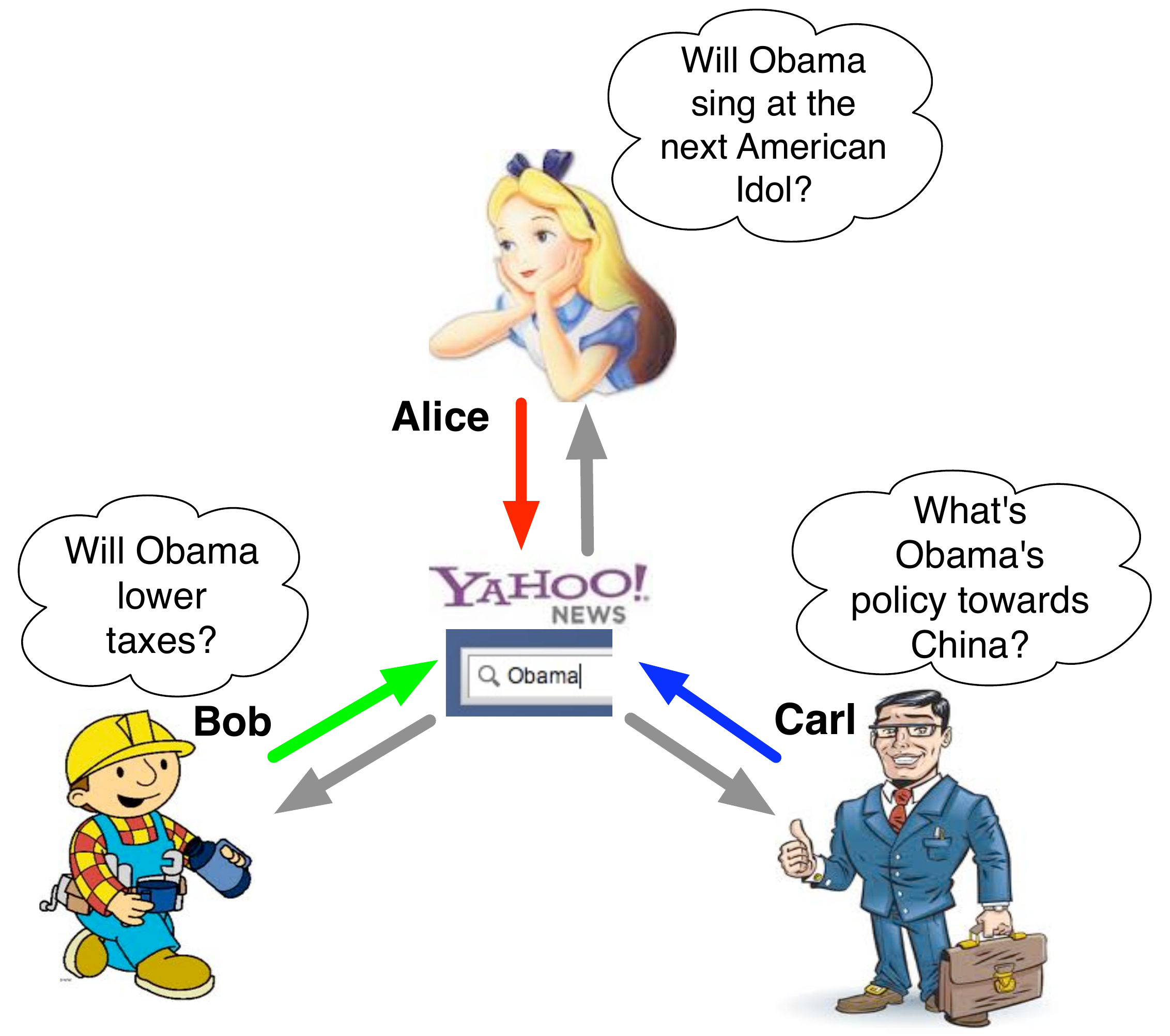}  % use this if you use "pdflatex"
	\caption{Different information need with the same short query}
	\label{fig:customized} % Fig.2
\end{figure}

However, users' information needs are difficult to gauge from individual search queries which are mostly short and succinct, and provide few details on an individual's personal preferences. A concrete example is shown in Fig.~\ref{fig:customized}, where Alice, Bob and Carl are interested in the latest news about \emph{President Obama} and submit the query ``obama" to Yahoo News. However, they each care about different topics related to ``President Obama'', represented by differently colored arrows. However, the search engine will not be able to capture such contextual information from the users' queries. The final ranking will thus be the same for all of the three users, represented by unified gray arrows.        

Considerable effort has been invested in providing personalized search and ranking services, but the prevalence of short search queries, in the absence of any other user-provided information, has long been a critical challenge for IR ranking algorithms.  Existing methods attempt to provide a more detailed assessment of users' interest by analyzing past user behavior, e.g. by means of the analysis of query logs~\cite{limam}, clickstream data~\cite{liu}, and users feedback~\cite{joachims}. Although promising results are obtained from the above methods, they only perform well when sufficiently large amount of user data is available and may raise significant privacy issues.

A promising direction in this domain has been the enrichment of short user queries with the users' community social context~\cite{xiaozhong}, which is based on the premise that information about the community that users are part of may help search engines to disambiguate particular facets of their information needs. Search results and rankings are thus geared not merely to respond to short, individual user queries themselves, but are enriched by information about the community that the individual user is part of. Search results and rankings are, in other words, ``communitized".

Thanks to the broad prevalence of social media, vast amounts of user-generated, real-time information is now open accessible online and can be used as a dynamic indicator of users' interest. Many efforts have therefore focused on mining users' interest from different types of social media content, such as Facebook~\cite{jin} and Twitter~\cite{wis} data. In particular, ~\cite{xiaozhong} developed a Community Interest Model to improve both web and news search rankings using blog data, and proved the efficacy of this approach by comparing its results with those produced by Google and Yahoo. However, blog data has two significant limitations: (1) it only represents the global community interest  and does not take into account the differences between geographical communities in the absence of adequate geo-location information, and (2) blog data does not respond well to rapid changes in user interest.

These limitations may be addressed by relying on data generated by Twitter, presently the most popular micro-blogging platform. Twitter data has a number of distinct advantages for those seeking community-enriched, dynamic information on news data. First, Twitter exposes users' real-time interest from their continuous stream of 140-character ``tweets''. Ten of million of Tweets are submitted on a daily basis by hundreds of millions of users. Second, Twitter provides explicit user geo-location data in its user profiles. Third, ~\cite{kwak} confirmed that most of the topics discussed on Twitter are actually headline news in media, which is appropriate given Twitter's design as a news sharing and dissemination service. In summary, the availability of large amount of Tweets, enriched with timestamps and users' geo-location information, as well the close connection of its content to the news media make Twitter a desirable indicator of users' real-time and localized interest towards news, which can be fully leveraged to improve news ranking based on community interest.

%[I'm not confident about this paragraph but I think it is necessary to mention this]. Actually, many main news search engines have already incorporated the region information into news ranking. However, it is only applied to the news that explicitly mention relevant region names. For example, for the Indiana local news, only the news that contains `Indianapolis', `Bloomington', etc., will appear as the top ranking results. Lots of news without explicit region names but also satisfy local users' interest is ignored.      
In this paper, we attempt to solve the above-mentioned problem of data sparsity by using \emph{dynamic community interest} gauged from tweets submitted from within a particular geographical community, defined as a US state. We show how such information can be leveraged to improve the rankings of news search results.
%We adopt this particular approach for two specific reasons: (1) much larger amounts of online data is available about for Twitter communities than can be obtained for individual users; (2) persons from the same geographical community are very likely to share similar interest, based on the theory of \emph{homophily}~\cite{birds}. In other words, we aim to provide \emph{communitized information rankings} that take into account a person's membership of a particular geographical community, in this case one of the 50 US states.

We propose a \emph{Community Tweets Voting Model (CTVM)}, and assess its effectiveness in re-ranking search results generated by Google News and Yahoo News on the basis of tweets collected from three US states, i.e., California (CA), New York (NY) and Texas (TX). We assess the quality of the various rankings by means of the \emph{Amazon Mechanical Turk(MTurk)}\footnote{https://www.mturk.com/mturk/welcome}. Our main findings show that CTVM can improve news ranking from Yahoo and Google for CA and NY, but does not seem to work well for TX.

%We address three questions. (1) \emph{whether community tweets can improve news ranking}. (1) \emph{whether tweets from one state can provide better ranking results than other two states.} (3) \emph{whether CTVM applies to all three states under study?}
%Our three research questions are:
%\begin{itemize}
%\item (1) Whether tweets can improve news ranking?
%\item (2) Whether tweets from one state can provide better ranking results than those from other two states?
%\item (3) Whether our voting model applies to all three states?
%\end{itemize}

\section{Community Tweets Voting Model}
\label{sec:overview}
We hypothesize that if the content of a news item is very similar to the tweets recently submitted by a particular community, it will be more in line with that community's interest, and therefore deserve a higher ranking in the generated search results for members of that community. Unlike other traditional ranking algorithms, we send queries to both Twitter and news search engines. For each news item in a particular search result set, we analyze the most recent tweets on that topic  from the particular geographical community. The tweets ``vote" to increase the news item's importance score, which is used to determine its optimal ranking.  The localized tweets are used to optimize the news ranking for each target community. 

Given a list of queries $\overrightarrow{\mathbf{Q}}=[q_1,...,q_r]$, $\overrightarrow{\mathbf{N}}_{q_r}^{k}=[N_1,...,N_k]$ represents top $k$ documents containing $q_r$ returned from news search engine, and $\overrightarrow{\mathbf{T}}_{q_r}^{s}=[T_1,...,T_m]$ represents all tweets containing $q_r$ collected from state $s$ on the same data when news results are extracted. A voting score vector $\overrightarrow{\mathbf{V}}_{q_r}^{s}=V(\overrightarrow{\mathbf{T}}_{q_r}^{s},\overrightarrow{\mathbf{N}}_{q_r}^{k})=[V_1,...,V_k]$ can be defined as:
\begin{equation}
V_j=Vote(\overrightarrow{\mathbf{T}}_{q_r}^{s} \to N_j)=\sum_{i=1}^{m}Sim(T_i,N_j),j=1,...,k
\label{eq:vote}
\end{equation}

\begin{figure*}[htbp]\centering
	\includegraphics[width=2\columnwidth]{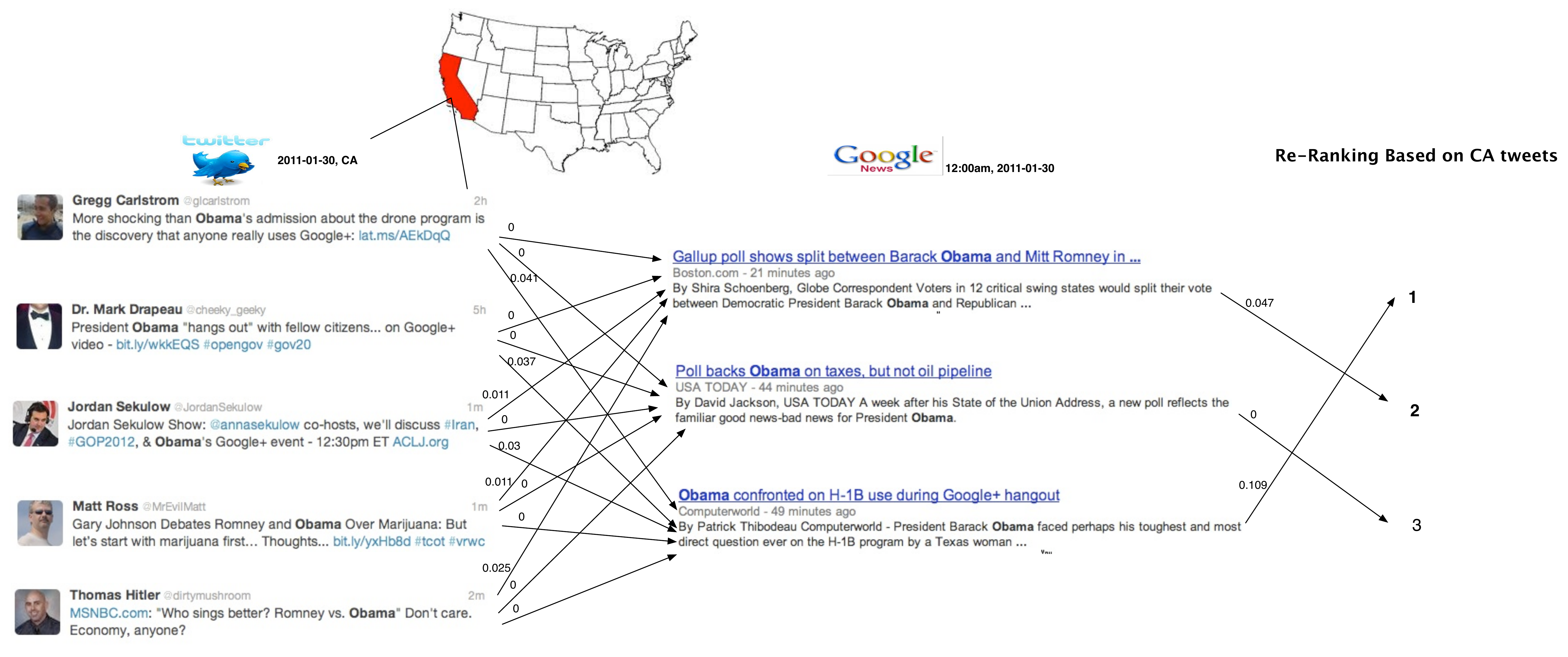}  % use this if you use "pdflatex"
	\caption{An example of CTVM modifying news item rankings using CA tweets for the query ``obama''.}
	\label{fig:vote} % Fig.2
\end{figure*}

In order to calculate $Sim(T_i,N_j)$, we define the vector space representation of $T_i$ and $N_j$. Due to the 140-characters space limit, a tweet generally contains very concise but topical words that can be considered as a ``short title''. Therefore, we compare the entire textual body of a tweet with the title of a news document, and define their vector representation as: $T_i=[w_T(t_1),...,w_T(t_h)]$ and $N_j=[w_N(t_1),...,w_N(t_h)]$ respectively, where $t_1,...,t_h$ is the common set of stemmed words shared by $T_i$ and $N_j$ after removing stop words and query words in $q_r$, $w_T(t_x)$ represents the term frequency of $t_x$ in $T_i$ and $w_N(t_x)$ represents the term frequency of $t_x$ in $N_j$. Therefore, the similarity score between $T_i$ and $N_j$ can be calculated as:

\begin{equation}
Sim(T_i,N_j)=\frac{\sum_{x=1}^{h}w_T(t_x)\cdot w_N(t_x)}{\sqrt{\sum_{x=1}^{h}w_{T}^{2}(t_x)}\cdot \sqrt{\sum_{x=1}^{h}w_{N}^{2}(t_x)}}
\label{eq:sim}
\end{equation}

An example of CTVM (purposely partly fictitious) is illustrated in Figure~\ref{fig:vote}. The top 3 returned documents for the query ``obama'' from Google News at 12:00pm on 2011-01-31 are shown in order from top to bottom. At the same time, 5 tweets matching the same query from CA are collected on 2011-01-31. The similarity scores of every pair of tweet and news document are calculated according to Equation~\ref{eq:sim}, and are shown on the arrows pointing from the tweets to the news documents. Subsequently, $\overrightarrow{\mathbf{V}}_{\text{obama}}^{\text{CA}}$ is calculated according to Equation~\ref{eq:vote}. The top 3 results are finally re-ranked accordingly. $\overrightarrow{\mathbf{V}}_{\text{obama}}^{\text{NY}}$ and $\overrightarrow{\mathbf{V}}_{\text{obama}}^{\text{TX}}$ can be calculated In a similar fashion. Four different rankings are provided for CA users: original search engine ranking (e.g.~Google), a localized tweets ranking $\overrightarrow{\mathbf{V}}_{\text{obama}}^{\text{CA}}$, and two non-localized tweets rankings $\overrightarrow{\mathbf{V}}_{\text{obama}}^{\text{NY}}$ and $\overrightarrow{\mathbf{V}}_{\text{obama}}^{\text{TX}}$. A similar process can be applied to other queries, other US states, and Yahoo News.

%
%\begin{table*}[htb]
%	\centering
%	\begin{tabular}{c|ccccccccccccccccccccccccccccccccc}
%	query&google&egypt&siri&tax&greece&election&nba&education&finance&kobe bryant&military&revolution&economy&vacation&insurance&obama&ncaa&cnn&christmas&iran&discount&britney spears&clinton&debt&republication&euro&lady gaga&lebron james&stock&nfl&health care&china&gay\\
%	\end{tabular}
%\end{table*}[htb]

\begin{table}[htb]
	\centering
	\scalebox{0.8}{
	\begin{tabular}{r|c|c|c|c|c|c|c|c|c} 
	\multirow{2}{*}{Query} 	 &\multicolumn{3}{|c|}{CA} 	     		&\multicolumn{3}{c|}{NY}				&\multicolumn{3}{c}{TX}\\ 
	\cline{2-10}
	
						&G&Y&T							&G&Y&T							&G&Y&T\\
	
	\hline
	\hline
	google				&20&20&2851						&20&20&1760						&&&\\
	egypt				&10&10&188						&&&								&&&\\	
	siri					&10&10&314						&&&								&&&\\
	tax					&60&60&1439						&60&60&1254						&30&30&850\\	
	greece/greek				&10&10&314						&&&								&&&\\
	election				&20&20&246						&20&20&190						&10&10&151\\
	nba					&70&70&1714						&80&80&1598						&20&20&1407\\
	education				&10&10&1090						&&&								&&&\\
	financial/finance		&10&10&1108						&&&								&&&\\
	kobe bryant			&80&80&1421						&90&90&1272						&30&30&960\\
	military/army			&20&20&1640						&&&								&20&20&1214\\
	revolution				&30&30&350						&30&30&281						&&&\\
	economy/economic		&60&60&984						&80&80&753						&30&30&960\\
	vacation				&20&20&889						&20&20&699						&&&\\
	insurance				&10&10&783						&&&								&&&\\
	obama				&50&50&2482						&70&70&1794						&20&20&1656\\
	ncaa					&40&40&481						&50&50&240						&&&\\
	cnn					&20&20&487						&20&20&468						&&&\\
	christmas				&50&50&25099					&70&70&19096					&40&40&20529\\
	iran					&20&20&323						&20&20&307						&10&10&177\\
	discount				&60&60&477						&60&60&1003						&50&50&289\\
	britney spears			&10&10&443						&&&								&&&\\
	clinton				&10&10&409						&&&								&&&\\
	debt					&10&10&473						&&&								&&&\\
	republication			&70&70&1728						&70&70&1320						&40&40&1290\\
	euro					&40&40&423						&50&50&496						&10&10&286\\
	lady gaga				&60&60&986						&60&60&1508						&30&30&386\\
	lebron james			&20&20&1082						&20&20&902						&10&10&484\\
	stock					&50&50&1492						&70&70&1461						&20&20&1103\\
	nfl					&60&60&2277						&60&60&1225						&40&40&1867\\
	health care			&30&30&468						&30&30&243						&20&20&280\\
	china				&30&30&1317						&40&40&945						&20&20&797\\
	gay/lesbian			&20&20&2887						&20&20&2214						&10&10&1835\\			
	\end{tabular}
	}
	\footnotetext[1]{\# of news results judged by Turkers form Google}
	\footnotetext[2]{\# of news results judged by Turkers from Yahoo}
	\footnotetext[3]{\# of tweets per day}			
	\caption{Data summary: $G$ and $Y$ indicate total \# of \textbf{G}oogle and \textbf{Y}ahoo news results evaluated by Amazon Turkers; $T$ indicates \# of tweets per day }
	\label{tab:data}
\end{table}

\section{Experiment}

\subsection{Data}
To test the performance of CTVM, four types of data were collected: test queries, tweets from the three states, daily ranking results from Google\&Yahoo search engines and users oriented interest judgement from MTurk.

\subsubsection{Test Queries Collection}
For our experiment, 50 initial test queries were selected for evaluation because they were known to be popular during the time period of evaluation. We singled out popular queries for two reasons. First, popular queries ensure that a reasonable quantity of matching tweets can be collected. Second, it is more likely that users understand popular queries well and can provide better judgements on the relevance of retrieved results for that query. Popular queries were manually identified by using \emph{Google Insights}\footnote{http://www.google.com/insights/search}. To increase the number of relevant tweets, some queries use different expressions of similar semantics, e.g., ``economics'' vs. ``economy'', ``gay'' vs. ``lesbian'', and ``army'' vs. ``military''.  

\subsubsection{Tweets Collection}
Daily tweets from CA, NY and TX were collected using Twitter streaming API for 50 selected queries from 2011-12-10 to 2011-12-24. Especially, the user location text returned from the API can be used to judge whether the tweet is from the three states, or not. If the full name or the capitalized two-letter abbreviation of the three states is found in the user location text, the corresponding tweet is collected. Finally, we collected 1,264,828 tweets from 250,549 CA users, 1,002,945 tweets from 195,637 NY users and 839,966 tweets from 161,948 TX users. 

\subsubsection{Ranking Results Collection}
Top 10 ranked retrieval results from Google News and Yahoo News were collected every day at 12:00pm from 2011-12-10 to 2011-12-24. The ranking position of each news document was stored along with news title, snippet, HTML content and date.

\subsubsection{User Judgement Collection}
We need real time users' judgements to serve as the ground truth to compare different ranking results for CTVM, Google and Yahoo. Since use's interest towards news is quite dynamic, we need to collect users' judgement results in near real-time. The evaluation task was setup by MTurk immediately after the news ranking lists were extracted from Google and Yahoo.

\begin{figure}[htbp]\centering
	\includegraphics[width=\columnwidth]{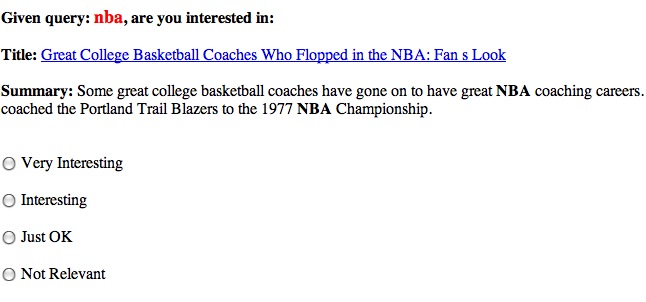}  % use this if you use "pdflatex"
	\caption{Amazon Mechanical Turk interface}
	\label{fig:turk} % Fig.2
\end{figure}
With instruction and examples, MTurk users (Turkers) were asked to provide real-time interest and relevance assessments towards a list of 20 news documents (10 each from Google or Yahoo) with respect to a specific query. For each query and for each day, the documents were shown in random order on one evaluation page, called a MTurk HIT. In the judgement process, for each retrieved document, the HIT presented the target query, the title with a hyperlink to the actual page and a snippet from the search engine. Turkes were required to choose one interest and relevant level from four choices: Very Interesting, Interesting, Just OK and Not Relevant.
A screenshot of one document judgement in one HIT for query ``nba" is shown in Figure~\ref{fig:turk}. To minimize the unbalanced distribution of the amount of Turkers over different queries due to potential individual bias, up to 5 different Turkers work on each HIT and up to 6 queries were put on MTurk every day.  
For any HIT, if fewer than three Turkers worked on it, the HIT was deleted from the database and wasn't used for evaluation. Each interest level is attached to a score: Very Interesting(3), Interesting(2), Just OK(1), Not Relevant(0). 
To make sure that Turkers are from the three selected states, we set up a pre-filteration process by checking the Turker's ip-address and mapping them into geo-lacation information using MaxMind GeoIP JavaScript\footnote{http://www.maxmind.com/app/}. The same HIT can be evaluated by Turkers from three states and the final judgement score of each HIT for each state is the mean of all Turkers' scores from the target state.
%\subsubsection{Data Summary}

Finally, there were 105 distinct Turkers that participated in this evaluation and a total of 5320 news documents were judged for 33 queries from 2011-12-10 to 2011-12-24. The rest 17 queries were removed because they were evaluated by less than three Turkers.

\subsubsection{Data Summary} 
Table~\ref{tab:data} shows the total number of news documents retrieved from Google and Yahoo and interest judgements via MTurk, along with the number of tweets collected per day, for all three states and all 33 queries. The blank entries indicate that no Turkers' judgements are received for certain queries and certain states. 
Obviously, the amount of CA Turkers and CA tweets dominate over the other two states, which corresponds to its 1st population size and Internet users amount\footnote{http://www.internetworldstats.com/unitedstates.htm} in US.

\subsection{Evaluation}
 The goal of evaluation is to compare four rankings mentioned in Section~\ref{sec:overview} for both Google News and Yahoo News: engine ranking, localized community tweets ranking and two non-localized community tweets rankings. \emph{Normalized Discounted Cumulative Gain (NDCG)}~\cite{jarvelin} is used to measure the effectiveness of certain ranking method towards a ranking list of news. The basic idea of NDCG is that a good ranking method always ranks relatively more relevant documents at higher positions. As we introduced in Section~\ref{sec:overview}, given a list of queries $\overrightarrow{\mathbf{Q}}=[q_1,...,q_r]$, $\overrightarrow{\mathbf{N}}_{q_r}^{k}=[N_1,...,N_k]$ represents top $k$ documents containing $q_i$ returned from news search engine. Let $R(q_i,N_j)$ be the relevance score assigned to document $N_j$ for query $q_i$, then
% we use NDCG@k to denote the ranking performance of top $k$ document and can be calculated as:
 \begin{equation}
NDCG@k(\overrightarrow{\mathbf{Q}})=\frac{1}{r}\sum_{i=1}^{r}Z_{ik}\sum_{j=1}^{k}\frac{2^{R(q_i,N_j)-1}}{log_2(1+i)} 
 \label{eq:ndcg}
 \end{equation} 
 where $Z_{ik}$ is a normalization factor calculated to make it so that a perfect ranking's NDCG@k for query $q_i$ is 1 and $R(q_i,N_j)$ is provided by Turkers from three states.

For each state, we calculated NDCG@3, NDCG@5, and NDCG@10 for both Google and Yahoo retrieval results and four ranking methods were compared: search engine ranking, CTVM ranking with local state tweets and CTVM ranking with two non-localized states tweets. The results for three states are shown in Table~\ref{tab:CA},~\ref{tab:NY} and ~\ref{tab:TX} respectively.
\begin{table}[htb]
	\centering
	\begin{tabular}{l || l l l l l} 
	Google News	      		&NDCG@3 	     		&NDCG@5		&NDCG@10\\ 
	\hline
	Google 				&\textbf{0.9031}	    	&\textbf{0.8621}	&0.8148\\ %\hline
	CTVM with CA 	 		&0.8930				&0.8432			&0.8168*\\ %\hline
	CTVM with NY			&0.8914				&0.8547			&0.8241*\\
	CTVM with TX			&0.8963				&0.8518			&\textbf{0.8293}*\\
	\hline
	\hline
	Yahoo News			&					&				&\\
	\hline
	Yahoo				&0.8801	    			&0.8387			&0.8101\\ %\hline
	CTVM with CA 	 		&\textbf{0.9156}*		&\textbf{0.8762}*	&\textbf{0.8370}*\\ %\hline
	CTVM with NY			&0.8992*				&0.8716*			&0.8309*\\
	CTVM with TX			&0.8950*				&0.8628*			&0.8254*\\
	\multicolumn{4}{l}{* denotes better than corresponding engine news ranking}\\
	\end{tabular}			
	\caption{Ranking performance comparison for CA. The ground truth comes from the CA Turkers. For both Google News and Yahoo News, the first line represents the engine ranking; The second line represents the localized (i.e., CA) tweets ranking and the rest two lines represent non-localized (i.e., NY, TX) tweets ranking. The rest two tables have the similar data layout.}
	\label{tab:CA}
\end{table}

\begin{table}[htb]
	\centering
	\begin{tabular}{l || l l l l l} 
	Google News	      		&NDCG@3 	     		&NDCG@5		&NDCG@10\\ 
	\hline
	Google 				&0.9020	    			&0.8628			&0.8375\\ %\hline
	CTVM with NY 	 		&0.9076*				&0.8628			&0.8412*\\ %\hline
	CTVM with CA			&\textbf{0.9182}*		&\textbf{0.8721}*	&\textbf{0.8436}*\\
	CTVM with TX			&0.8853				&0.8588			&0.8384*\\
	\hline
	\hline
	Yahoo News			&					&				&\\
	\hline
	Yahoo				&0.8869	    			&0.8567			&0.8314\\ %\hline
	CTVM with NY 	 		&\textbf{0.9255}*		&\textbf{0.8874}*	&\textbf{0.8639}*\\ %\hline
	CTVM with CA			&0.9070*				&0.8703*			&0.8604*\\
	CTVM with TX			&0.8990*				&0.8671*			&0.8449*\\
	\multicolumn{4}{l}{* denotes better than corresponding engine news ranking}\\
	\end{tabular}			
	\caption{Ranking performance comparison for NY}
	\label{tab:NY}
\end{table}

\begin{table}[htb]
	\centering
	\begin{tabular}{l || l l l l l} 
	Google News	      		&$NDCG@3$ 	     		&NDCG@5		&$NDCG@10$\\ 
	\hline
	Google 				&\textbf{0.8630}	    	&\textbf{0.8289}	&0.7976\\ %\hline
	CTVM with TX 	 		&0.8203				&0.7874			&0.7859\\ %\hline
	CTVM with CA			&0.8535				&0.8252			&\textbf{0.8067}*\\
	CTVM with NY			&0.8355				&0.8058			&0.7875\\
	\hline
	\hline
	Yahoo News			&					&				&\\
	\hline
	Yahoo				&\textbf{0.8199}   		&\textbf{0.7865}	&\textbf{0.7754}\\ %\hline
	CTVM with TX 	 		&0.7863				&0.7448			&0.7420\\ %\hline
	CTVM with CA			&0.7728				&0.7461			&0.7497\\
	CTVM with NY			&0.8046				&0.7682			&0.7649\\
	\multicolumn{4}{l}{* denotes better than corresponding engine news ranking}\\
	\end{tabular}			
	\caption{Ranking performance comparison for TX}
	\label{tab:TX}
\end{table}
 
The most illustrative observation about Yahoo news ranking in both Table~\ref{tab:CA} and Table~\ref{tab:NY} is that CTVM with localized tweets perform the best in re-ranking the news documents in CA and NY, especially for top 3 relevant news, which has two implications: 1, CTVM is very effective in improving news ranking for Yahoo; 2, compared with non-localized information, localized tweets can further enhance the 3 most relevant news ranking for Yahoo, by incorporating local community interest. Plus, CTVM with tweets from any of the three states outperforms Yahoo, regardless of NDCG@3, NDCG@5 or NDCG@10, implying that adding users' interest (even not localized) to the news ranking is always good for Yahoo. 

The improvement of CTVM to Google news ranking is also spotted in Table~\ref{tab:CA} and Table~\ref{tab:NY}, although not as evident as Yahoo. Specifically, CTVM with any of the three states performs better than Google in top 10 news ranking but not always as good as Google in top 3 and 5 news ranking. It indicates that Google news ranking itself is a relatively robust ranking method which may have already utilized users' interest information more or less, especially for the top 3 news. In addition, the observation that CTVM with localized tweets does not perform any better than non-localized tweets implies that the news document selected and indexed by Google are generally universally interesting which minimizes the regional difference.

By contrast, Table~\ref{tab:TX} shows that CTVM does not perform well for TX users, because the original rankings are generally better than those generated by CTVM. We speculate that two possible factors may be responsible for this observation. First, both the amount of  Amazon Mechanical Turk workers and the number of Tweets are lowest for TX among the three states we investigated (see Table~\ref{tab:data}). Consequently, TX data may be less reliable than that of the two other states. Second, users' interest mined from TX tweets may be inconsistent with TX users' interest in headline news, violating the main assumption behind the proposed CTVM. To investigate this possibility, we selected the query ``china'' and manually examined a sample of tweets that contained the term ``china'' for all three states. We found that CA and NY tweets seem to be mostly about the economy and politics of China, and contain hyperlinks that point to news sites. By contrast, TX tweets seem to be mostly about Chinese products and artifacts, and as a result contain hyperlinks that point to general websites (e.g. Amazon), instead of news sites. Further investigation is required to determine whether the mentioned reasoned can indeed explain the lesser performance of CTVM for TX users, but it is clear that certain geographical communities may have characteristics that are at odds with the basic assumption underlying CTVM.

\subsection{Discussion}
The evaluation shows that CTVM achieved good performance in providing communitized, real-time news ranking for CA and NY users, but not for TX users. In addition, CTVM in particular improves Yahoo news rankings.  

We now propose an application scenario for CTVM. The low cost and barriers to implementation of the CTVM ranking method can benefit a large number of local news provider; it is easy to integrate CTVM into any search engines without the requirement to obtain large-scale network topology, query log, feedback, or clickstream data. Rather, search engines that adopts CTVM need only to acquire daily tweets from some selected states (or cities, countries). When users enter their search queries, the search engine can conveniently retrieve their {IP}-addresses, match them to the stored community model for the geographical location, and modify the voting scores of the top $k$ news documents using real-time tweets from the users' geo-location. The search engine can offer users the options of whether to activate CTVM re-ranking, choose their preferred $k$ value, and use either localized or non-localized tweets. If the amount of tweets from a particular location is insufficient, tweets from adjacent locations can still be used to provide augmented rankings. It is furthermore straightforward to extend CTVM with the analysis of personal user data, such as query log and session mining, clickstream analysis and users feedback analysis.

\section{Related Work}
\subsection{Ranking}
The development and refinement of ranking mechanism has been always at the core of IR research. Content-based ranking and linkage-based ranking are two classical models. Content-based methods rank documents according to how their content matches a given search query, and may rely on vector space models~\cite{salton} and language models~\cite{lv}. Linkage-based methods rank documents according to their position in the topology of hyperlink networks, e.g.~PageRank~\cite{ilprints} and HITS~\cite{kleinberg}. These methods however do not take into account users interest that are not expressed in search queries, document content or network topology.

Recently, researchers have explored applications of online behavior data, query sessions, logs~\cite{limam}, clickstream data~\cite{liu}, and users feedback~\cite{joachims} data, to generate personalized search rankings. Although these have been proven to be effective, they require large amounts of user behavioral data which can be difficult to obtain and manage. \cite{xiaozhong} attempts to solve the data sparsity problem by substituting personal data with community data on the assumption that people from the same community share similar interest. His work, however, relies on global community interest. As an extension, our work partitions different communities by geo-location to provide localized rankings.  

\subsection{Twitter data analytics}
Several studies has leveraged the collective behavior of Twitter users to gain insight into a number of real-life phenomena. Analysis of tweet content has shown correlations between users' global moods and important worldwide events\\\cite{johan}. Twitter can be also used to predict stock market fluctuations~\cite{stock} and earthquakes~\cite{sakaki}. 

Since~\cite{kwak} has confirmed the close relation of Twitter to headline news, we have seen numerous explorations of Twitter data to news analytics. \cite{phuvipadawat} developed an system to detect and track breaking news in real-time, and~\cite{wis} modeled user's interest from Twitter and provided personalized news for Twitter users. However, to the best of our knowledge few studies have used Twitter data to optimize and augment search engine rankings of news items generated by traditional search engines such as Google and Yahoo.

\section{Conclusion}
In this paper, we propose the CTVM to re-rank news search results retrieved from Google and Yahoo using an analysis of tweets from three states: CA, NY and TX, based on the assumption that  assessments of users' interest in news can be augmented on the basis of information about their geographical community. We validate our results by obtaining ground-truth assessment of ranking quality from Amazon's Mechanical Turk. Preliminary experimental results show that CTVM outperforms Yahoo in its top 3, 5, 10 news document rankings and outperforms Google in its top 10 news documents rankings. This is the case for both CA and NY communities. In addition, in CA and NY, CTVM using local tweets performs better than using non-local tweets for Yahoo news ranking. This implies that users' regional preferences make a greater difference for Yahoo news rankings than Google's. TX is the exception on all CTVM performance indicators. We hypothesize that this is either caused by insufficient ranking evaluations and tweets from TX, or the fact that TX tweets do not match the news interest of TX residents.

In spite of these promising results, numerous issues merit further investigation. First, we propose to further explore CTVM's poor performance for the TX community which may result from interesting regional and social variations. Second, the CTVM could employ Named Entity Recognition or other NLP tools to determine semantic instead of word similarity to adjust voting scores. These methods need to acknowledge the real-time, temporal dynamics of changing users' interest. Third, our assessment relied on the average performance of CTVM over all queries and did not consider the differences between queries in terms of their general subject matter, e.g.~politics, science, entertainment, and celebrity news, and their different temporal properties (i.e.~hypes and fads vs. long-standing discussions). Finally, CTVM may be extended beyond location-based communities to include other demographic factors, such as gender, age, and even mood \cite{mood:bollen2011} which can equally be used to demarcate online communities, and may in fact provide a more reliable definition of news-relevant communities.

%ACKNOWLEDGMENTS are optional
%\section{Acknowledgments}
%This section is optional; it is a location for you
%to acknowledge grants, funding, editing assistance and
%what have you.  In the present case, for example, the
%authors would like to thank Gerald Murray of ACM for
%his help in codifying this \textit{Author's Guide}
%and the \textbf{.cls} and \textbf{.tex} files that it describes.
\bibliographystyle{abbrv}
\bibliography{refs}  % sigproc.bib is the name of the Bibliography in this case

\begin{thebibliography}{10}

\bibitem{wis}
F.~Abel, Q.~Gao, G.-J. Houben, and K.~Tao.
\newblock {Analyzing Temporal Dynamics in Twitter Profiles for Personalized
  Recommendations in the Social Web}.
\newblock In {\em Proceedings of ACM WebSci '11, 3rd International Conference
  on Web Science, Koblenz, Germany}, 2011.

\bibitem{mood:bollen2011}
J.~Bollen, B.~Gon\c{c}alves, R.~GuangChen, and H.~Mao.
\newblock Happiness is assortative in online social networks.
\newblock {\em ALife}, 17(3):237--251, 2011.

\bibitem{johan}
J.~Bollen, H.~Mao, and A.~Pepe.
\newblock Modeling public mood and emotion: Twitter sentiment and
  socio-economic phenomena.
\newblock In {\em ICWSM}, 2011.

\bibitem{stock}
J.~Bollen, H.~Mao, and X.-J. Zeng.
\newblock Twitter mood predicts the stock market.
\newblock {\em J. Comput. Science}, 2(1):1--8, 2011.

\bibitem{jarvelin}
K.~J\"{a}rvelin and J.~Kek\"{a}l\"{a}inen.
\newblock Cumulated gain-based evaluation of ir techniques.
\newblock {\em ACM Trans. Inf. Syst.}, 20:422--446, October 2002.

\bibitem{jin}
X.~Jin, C.~Wang, J.~Luo, X.~Yu, and J.~Han.
\newblock Likeminer: a system for mining the power of 'like' in social media
  networks.
\newblock In {\em Proceedings of the 17th ACM SIGKDD international conference
  on Knowledge discovery and data mining}, KDD '11, 2011.

\bibitem{joachims}
T.~Joachims and F.~Radlinski.
\newblock Search engines that learn from implicit feedback.
\newblock {\em Computer}, 40:34--40, August 2007.

\bibitem{kleinberg}
J.~M. Kleinberg.
\newblock Authoritative sources in a hyperlinked environment.
\newblock {\em J. ACM}, 46:604--632, September 1999.

\bibitem{kwak}
H.~Kwak, C.~Lee, H.~Park, and S.~Moon.
\newblock What is twitter, a social network or a news media?
\newblock In {\em Proceedings of the 19th international conference on World
  wide web}, WWW '10, 2010.

\bibitem{limam}
L.~Limam, D.~Coquil, H.~Kosch, and L.~Brunie.
\newblock Extracting user interests from search query logs: A clustering
  approach.
\newblock In {\em Proceedings of the 2010 Workshops on Database and Expert
  Systems Applications}, DEXA '10, 2010.

\bibitem{liu}
J.~Liu, P.~Dolan, and E.~R. Pedersen.
\newblock Personalized news recommendation based on click behavior.
\newblock In {\em Proceedings of the 15th international conference on
  Intelligent user interfaces}, IUI '10, 2010.

\bibitem{xiaozhong}
X.~Liu and V.~von Brzeski.
\newblock Computational community interest for ranking.
\newblock In {\em Proceedings of the 18th ACM conference on Information and
  knowledge management}, CIKM '09, 2009.

\bibitem{lv}
Y.~Lv and C.~Zhai.
\newblock Positional language models for information retrieval.
\newblock In {\em Proceedings of the 32nd international ACM SIGIR conference on
  Research and development in information retrieval}, SIGIR '09, 2009.

\bibitem{ilprints}
L.~Page, S.~Brin, R.~Motwani, and T.~Winograd.
\newblock The pagerank citation ranking: Bringing order to the web.
\newblock Technical report, Stanford InfoLab, November 1999.

\bibitem{phuvipadawat}
S.~Phuvipadawat and T.~Murata.
\newblock Breaking news detection and tracking in twitter.
\newblock In {\em Proceedings of the 2010 IEEE/WIC/ACM International Conference
  on Web Intelligence and Intelligent Agent Technology - Volume 03}, WI-IAT
  '10, 2010.

\bibitem{sakaki}
T.~Sakaki, M.~Okazaki, and Y.~Matsuo.
\newblock Earthquake shakes twitter users: real-time event detection by social
  sensors.
\newblock In {\em Proceedings of the 19th international conference on World
  wide web}, WWW '10, 2010.

\bibitem{salton}
G.~Salton, A.~Wong, and C.~S. Yang.
\newblock A vector space model for automatic indexing.
\newblock {\em Commun. ACM}, 18:613--620, November 1975.

\end{thebibliography}
\end{document}